\documentclass[12pt,tightenlines,onecolumn,preprintnumbers,
nofootinbib,amsmath,amssymb,prd,showpacs]{revtex4}
\usepackage{colordvi}
\input colordvi
\usepackage{color}
\usepackage{graphicx}%
\usepackage{amsmath}
\usepackage{amssymb}
\usepackage{bm}
\usepackage{enumerate}

\newcommand{\nn}{\nonumber}

\newcommand{\beq}{\begin{equation}}
\newcommand{\eeq}{\end{equation}}
\newcommand{\beqa}{\begin{eqnarray}}
\newcommand{\eeqa}{\end{eqnarray}}
\newcommand{\bseq}{\begin{subequations}}
\newcommand{\eseq}{\end{subequations}}

\begin{document}
\title{String or branelike solutions in four-dimensional Einstein gravity
in the presence of cosmological constant}
\author{Youngone Lee${}^{a, ~d}$}
\email{ youngone@hanyang.ac.kr}
\author{Gungwon Kang \footnote{Corresponding authors: Gungwon Kang and Hyeong-Chan Kim} ${}^b$}
\email{gwkang@kisti.re.kr }
\author{Hyeong-Chan Kim${}^c$}
\email{hckim@cjnu.ac.kr}
\author{Jungjai Lee${}^a$}
\email{jjlee@daejin.ac.kr}
\affiliation{${}^{a}$ Department of Physics and Institute of Basic Sciences, Daejin University, Pocheon, Gyeonggi 487-711, Korea}
\affiliation{${}^b$ Korea Institute of Science and Technology
Information (KISTI), 334 Gwahak-ro, Yuseong-gu, Daejeon 305-806,
Korea}
\affiliation{${}^c$ School of Liberal Arts and Sciences, Chungju National
University, Chungju 380-702, Korea}
\affiliation{${}^d$ Research Institute for Natural Sciences, Hanyang University, Seoul 133-791, Korea}
\bigskip

\begin{abstract}
\bigskip
We investigate string or branelike solutions for four-dimensional
vacuum Einstein equations in the presence of cosmological constant.
For the case of negative cosmological constant, the Banados-Teitelboim-Zanelli black string is the only warped stringlike solution. The general solutions for
nonwarped branelike configurations are found and they are
characterized by the Arnowitt-Deser-Misner mass density and two tensions.
Interestingly, the sum of these tensions is equal to the minus of
the mass density. Other than the well-known black string and
soliton spacetimes, all the static solutions possess naked singularities.
The time-dependent solutions can be regarded as the anti-de Sitter extension of
the well-known Kasner solutions. The speciality of those static
regular solutions and the implication of singular solutions are also
discussed in the context of cylindrical matter collapse.
For the case of positive cosmological constant, the Kasner-de Sitter
spacetime appears as time-dependent solutions and all static
solutions are found to be naked singular.
\end{abstract}

\date{\today}
\maketitle

\section{Introduction}

One may define hypercylindrical spacetimes as spacetime configurations having translational symmetries along some spatial directions. Black $p$-brane solutions have such configurations and have been used in various studies in string theory.
In the context of the extra-dimensional models, a black hole observed in our four-dimensional universe could be a part of the full dimensional black string or brane configuration rather than a part of the full dimensional black hole confined around our universe. Gregory and Laflamme found that black string or $p$-brane backgrounds are generically unstable on the small perturbations in  the string or brane directions.
However, some black $p$-brane configurations such as Banados-Teitelboim-Zanelli (BTZ) black string show absence of the Gregory-Laflamme instability. The origin and the evolution of such instability have not been fully understood as yet. Therefore, it is very important to know the full solution space for string or brane configurations.

In general relativity, hypercylindrical static vacuum solutions in five spacetime dimensions were found repeatedly by several authors~\cite{Kramer,Chodos:1980df,Lee:2006jx}.
Initially, these solutions were investigated in the context of the Kaluza-Klein dimensional reduction, including the interpretation of integral constants as mass and scalar charge.
The correct interpretation of the integral constants as ADM mass density and gravitational tension and the study on the geometrical properties were given recently in Refs.~\cite{Lee:2006jx,CKKL}. The extension to spacetime dimensions higher than five and the inclusion of dilatonic scalar and antisymmetric form fields are considered in Refs.~\cite{kang,camps} and in Refs.~\cite{Stringtheory}, respectively. (See also references therein.) Stationary stringlike solutions having a momentum flow along the string direction were found by Chodos and Detweiler~\cite{Chodos:1980df}, and their geometrical properties were analyzed by Kim and Lee~\cite{kl} in detail.
Gravitational energy conditions on Arnowitt-Deser-Misner (ADM) mass and momentum densities and tension have also been analyzed recently~\cite{Kim:2010nz}, with presenting classification of solutions under boost transformations.

In this paper we search for the same type of solutions in the presence of cosmological constant. The inclusion of cosmological constant may change the behaviors of spacetime solutions drastically. For instance, the spacetime becomes asymptotically nonflat.
The Gregory-Laflamme instability~\cite{GL} generically occurring for black string or brane solutions disappears for a certain class of such solutions with a negative cosmological constant~\cite{Hirayama:2001bi}. Furthermore, in the presence of a positive cosmological constant, it was shown that the cosmological anisotropy will be exponentially washed away if the matter fields satisfy the dominant energy condition except for special cases in Bianchi IX~\cite{Wald}. Note also that our universe has a nonvanishing positive cosmological constant.
Therefore, it is very interesting to see how string or branelike solutions behave differently in the presence of cosmological constant.

At the present paper, we consider the case of four-dimensional spacetime only. In the absence of cosmological constant, black string solutions cannot exist in four dimensions~\cite{Horowitz:1991cd}, but start to appear from five dimensions. In the presence of cosmological constant, however, black string or brane solutions such as BTZ black string and anti-de Sitter (AdS) $C$-metrics in Ref.~\cite{Emparan:1999fd} exist even in four dimensions.
Therefore, instead of considering general higher dimensional spacetime solutions, we search for a wider class of string or branelike configurations in four dimensions for simplicity.
In addition, we also examine whether or not a wider class of the Kasner solutions with cosmological constant~\cite{Kasner-deSitter} exist.

In Sec.~\ref{warped}, we consider a class of stringlike solutions that are warped along a spatial direction. It is shown that the BTZ black string is indeed the unique nontrivial stringlike solution. In Sec.~\ref{solution}, general stringlike solutions without warping factor are considered. Static and time-dependent solutions are presented for the cases of negative and positive cosmological constants. In Secs.~\ref{-L} and~\ref{+L}, the causal structures and geometrical properties of those solutions are summarized.
Conclusions are followed in Sec.~\ref{discussion} and physical implications are also discussed.

\section{Warped solutions}
\label{warped}

The action of the Einstein gravity with cosmological constant in four dimensions is given by
\begin{eqnarray}
S=\frac{1}{16\pi G}  \int d^4 x \sqrt{-g}~(R-2\Lambda).
\end{eqnarray}
Here we are interested in searching for stringlike solutions as general as possible. Let us first summarize some results on the stringlike solutions for the case of five-dimensional Einstein gravity without cosmological constant ({\it i.e.}, $\Lambda = 0$). Consider hypercylindrical uniform static vacuum solutions in five dimensions whose metric forms are given by
 \beq
ds^2 = -F(\rho)dt^2 +G(\rho)\left( d\rho^2 +\rho^2 d\Omega_2^2
\right) +d\eta^2
 \label{metric}
 \eeq
in the isentropic coordinate system. We find that the only solution is the Schwarzschild black string where
 \beq
F(\rho)=\left({1-K/\rho \over 1+K/\rho} \right)^2, \qquad G(\rho)=
\left( 1+{K\over \rho}\right)^4 .
 \label{Schmetric}
 \eeq
This solution appears to be characterized by a single parameter, namely, the ADM mass density $M (= 2K)$. However, we expect that there should be another physical quantity ({\it i.e.}, the ADM tension) associated with the translational symmetry along the $\eta$ direction in this metric. Actually, the above solution turns out to be a special case where the ADM tension is given by $\tau = M/2 = K$. More general stringlike solutions having arbitrary values of tension were indeed obtained by allowing the metric component $g_{\eta\eta}$ arbitrary, {\it e.g.}, $g_{\eta\eta}=1 \rightarrow Z(\rho)$, and the solutions are~\cite{Lee:2006jx,CKKL}
 \beq
F = \left| \frac{1-\frac{K}{\rho}}{1+\frac{K}{\rho}}
\right|^{\frac{2(2 -a)}{\sqrt{3(a^2 -a +1)}}} = Z^{(2-a)/(2a-1)},
\qquad  G = \left( 1+\frac{K}{\rho} \right)^4 \left|
\frac{1+\frac{K}{\rho}}{1-\frac{K}{\rho}}
\right|^{\frac{2(a+1)}{\sqrt{3(a^2 -a +1)}} -2} .
 \eeq
These solutions are characterized by two parameters ({\it e.g.}, mass density and tension) and are still asymptotically flat at $\eta= {\rm constant}$ surfaces. However, most of them have naked singularity at $\rho=K$ surface and the only regular solutions occur when the tension to mass ratio
($\tau/M \equiv a$) is $a=1/2, 2$, which correspond to the Schwarzschild black string and the Kaluza-Klein bubble, respectively.

Now let us consider what happens if a cosmological constant is present. For simplicity we consider four dimensions. In four-dimensional Einstein gravity with a negative cosmological constant ($\Lambda < 0$), the well-known black string metric is the BTZ solution given by
 \beq
ds^2 = \frac{-3/\Lambda}{\sin^2 \eta} \left[ -\left( -M +r^2 \right)
dt^2 +\frac{dr^2}{-M +r^2} +r^2d\theta^2 +d\eta^2 \right] .
\label{BTZBS}
 \eeq
Note that this solution is asymptotically $AdS_4$ at $r \rightarrow \infty$. Note also that the string direction is not uniform, but warped. In order to find a more general class of
stringlike static vacuum solutions we consider the following metric ansatz, similarly to the case of the Schwarzschild black string above,
 \beq
ds^2 = H^{-2}(\eta,\rho) \left[ -F(\rho)dt^2 +G(\rho)\left( d\rho^2
+\rho^2 d\theta^2 \right) +Z(\rho) d\eta^2 \right] .
 \label{WarpM}
 \eeq
Note here that the $\rho$ dependence is allowed on the warping factor in addition to the arbitrary $\rho$-dependence in the $g_{\eta\eta}$ component.

The $(\rho,\eta)$ component of the field equations gives
 \beq
\frac{\partial}{\partial \eta}\left[2 \frac{\partial H(\eta,\rho)}{\partial \rho}
Z(\rho) - H(\eta,\rho) \frac{\partial
Z(\rho)}{\partial \rho}\right] = 0.
 \label{rhoeta}
 \eeq
Note first that this equation is easily solved if the warping factor is a function of $\rho$ only, {\it i.e.}, $H(\eta, \rho) = H(\rho)$. For such case that the geometry is not warped in the $\eta$-direction, we can set $H = 1$ by redefining the functions $F$, $G$ and $Z$, and it will be considered separately in the next section.

Now, if the geometry is warped in the $\eta$ direction, {\it i.e.}, $\partial_{\eta} H \not= 0$, Eq.~(\ref{rhoeta}) gives
 \beq
H(\eta,\rho) = \sqrt{Z(\rho)} \left[ h(\eta) +g(\rho) \right] .
 \eeq
Thus, we can set
 \beq
Z(\rho) = 1, \qquad  H(\eta,\rho) = h(\eta) +g(\rho)
 \eeq
by redefining the functions $F$ and $G$ in Eq.~(\ref{WarpM}). We first consider the case that the warping factor is independent of the $\rho$ coordinate, {\it e.g.}, $g(\rho) = 0$. Then, solving the rest of the field equations gives
 \beq
F = C_1 \tan^2 \sqrt{M}\ln (\rho/\rho_0) , \quad G =
\frac{-M/\Lambda_3}{\rho^2 \cos^2 \sqrt{M}\ln (\rho/\rho_0)} , \quad H
= C_2 e^{\sqrt{\Lambda_3}\eta} +C_3 e^{-\sqrt{\Lambda_3}\eta} .
 \label{warpedsols}
 \eeq
Here the integral constants are $C_1, C_2, C_3, \rho_0, M$ and $\Lambda_3$ with $C_2 C_3 = \Lambda/(12 \Lambda_3)$, and they must be chosen suitably to make the metric functions real.

i) $\Lambda < 0$; In this case the spacetime is asymptotically $AdS_4$. The warping function in Eq.~(\ref{warpedsols}) can be reexpressed as
 \begin{equation}
H = \left\{
  \begin{array}{ll}
      \sqrt{\Lambda/3\Lambda_3} \sin \sqrt{-\Lambda_3} (\eta -\eta_0)
& \hbox{\qquad {\rm for} \quad $\Lambda_3 < 0$.} \\ 
    \sqrt{-\Lambda/3\Lambda_3} \sinh \sqrt{\Lambda_3} (\eta -\eta_0)
& \hbox{\qquad {\rm for} \quad $\Lambda_3 > 0$,} \\ \label{W2}
    \sqrt{-\Lambda/3} (\eta -\eta_0)
& \hbox{\qquad {\rm for} \quad $\Lambda_3 = 0$,} 
  \end{array}
\right.
\end{equation}
The integral constants are restricted accordingly to the sign of $\Lambda_3$. That is, for $\Lambda_3 < 0$, $M, C_1 > 0$ and $C_2 = C_3^*$ with $|C_2| = \sqrt{\Lambda/(12 \Lambda_3)}$, and the metric is written as
 \beq
ds^2 = \frac{3\Lambda_3/\Lambda}{\sin^2 \sqrt{-\Lambda_3}\eta}
  \left[ -\tan^2 \sqrt{M}\ln \rho \, dt^2 +\frac{M/(-\Lambda_3)}{\rho^2 \cos^2 \sqrt{M}\ln \rho} \left( d\rho^2 +\rho^2 d\theta^2 \right) +d\eta^2 \right] .
 \label{Warped-L3}
 \eeq
Here $C_1$ is absorbed by rescaling the $t$-coordinate, and $\eta -\eta_0 \rightarrow \eta$ and $\rho/\rho_0 \rightarrow \rho$ are used. For $\Lambda_3 > 0$, $M, C_1 < 0$, and the metric is
 \beq
ds^2 = \frac{3\Lambda_3/(-\Lambda)}{\sinh^2 \sqrt{\Lambda_3}\eta}
  \left[ -\tanh^2 \sqrt{\bar{M}}\ln \rho \, dt^2 +\frac{\bar{M}/\Lambda_3}{\rho^2 \cosh^2 \sqrt{\bar{M}}\ln \rho} \left( d\rho^2 +\rho^2 d\theta^2 \right) +d\eta^2 \right] .
 \label{Warped+L3}
 \eeq
Here $\bar{M} = -M > 0$. When $\Lambda_3  =0$, we take $M \rightarrow 0$ and $C_1 \rightarrow \infty$ with $-M/\Lambda_3 \equiv G_0 \, (> 0)$ and $C_1 M \equiv F_0 \, (> 0)$ fixed. Hence one finds
 \beq
ds^2 = \frac{-3/\Lambda}{\eta^2}
  \left[ -\ln^2 \rho \, dt^2 +\frac{1}{\rho^2} \left( d\rho^2 +\rho^2 d\theta^2 \right) +d\eta^2 \right] .
 \label{Warped0L3}
 \eeq
Here $F_0$ and $G_0$ are absorbed by rescaling the $t$ and $\eta$ coordinates. Defining $\bar{r} = \ln \rho$, it may be rewritten as
 \beq
ds^2 = \frac{-3/\Lambda}{\eta^2} \left( - \bar{r}^2 dt^2 +d\bar{r}^2
+d\theta^2 +d\eta^2 \right) .
 \label{Warped0L3Rindler}
 \eeq
In order to see what these solutions look like geometrically, we can reexpress them in a more conventional form by introducing the areal radial coordinate, {\it e.g.}, $r^2 = \rho^2 G$,~\footnote{It should be point out that this coordinate transformation is not defined well for the case of $\Lambda_3 = 0$. However, the metric in Eq.~(\ref{WarpedsolsAreal}) is well defined in the limit of $\Lambda_3 \rightarrow 0$ with $-M >0$, and describes the pure $AdS_4$ geometry.}
 \beq
ds^2 = \frac{1}{H^2(\eta)} \left[ - (-M -\Lambda_3 r^2)
dt^2 + \frac{dr^2}{-M -\Lambda_3 r^2} +r^2 d\theta^2 +d\eta^2 \right] .
 \label{WarpedsolsAreal}
 \eeq
Here the constant $C_1$ is absorbed by rescaling the $t$ coordinate. The signs of $\Lambda_3$ and $M$ and the corresponding form of the function $H$ are as explained above. Note that the value of $\Lambda_3$ can be set to $-1, 0, +1$ by rescaling the coordinates according to its sign. Therefore, we see that the case of $\Lambda_3 < 0$ is the BTZ black string written in Eq.~(\ref{BTZBS}), and that the cases of $\Lambda_3 > 0$ and $\Lambda_3= 0$ are nothing but $dS_3$ (de Sitter) and $M_3$ (flat) foliations of the same pure $AdS_4$ spacetime, respectively.

ii) $\Lambda > 0$; In this case the spacetime is asymptotically $dS_4$.
The metric can be rewritten as in Eq.~(\ref{WarpedsolsAreal}) as well. However, the case of $\Lambda_3<0$ is not allowed since the function $H$ cannot be real valued ({\it e.g.,} $|C_2|^2= \Lambda/(12 \Lambda_3)$ cannot be positive). Therefore, the solutions are nothing but $dS_3$ and $M_3$ foliations of the same $dS_4$ spacetime.

Now let us consider the case that the geometry is warped not only along the $\eta$ direction, but also along the $\rho$ direction ({\it i.e.}, $g \neq {\rm constant}$). By solving all other field equations, we finally obtain
 \beq
H = \sqrt{(-\Lambda/3) C} \sin \frac{\eta - \eta_0}{\sqrt{C}} + \frac{A}{D + \ln \rho} ,
\quad  F = \frac{B}{(\ln \rho + D)^2}  , \quad G = \frac{C}{\rho^2 (\ln \rho + D)^2}.
 \eeq
Here we assumed $B, C > 0$. By defining $r = \ln \rho +D$ and rescaling $\sqrt{B/C} t \rightarrow t$, $(\eta -\eta_0)/\sqrt{C} \rightarrow \eta$, $A/\sqrt{C} \rightarrow A$, the metric in Eq.~(\ref{WarpM}) can be expressed as~\footnote{We point out that the above metric becomes identical to the four-dimensional analogy of the solution found in Ref.~\cite{Kim:1999ja} if we perform triple Wick rotations; $r = i \tau$, $t = i x$ and $\eta = i \sqrt{-\Lambda/3} b_0 y$.}
 \beq
ds^2 = \frac{1}{(\sqrt{-\Lambda/3} r \sin \eta +A)^2} \left( -dt^2 +dr^2 +d\theta^2 +r^2 d\eta^2 \right) .
 \eeq
This metric looks like a new solution, but it turns out to be the pure $AdS_4$.
Defining $x_3 = r \sin \eta$, $x_1 = r \cos \eta$ and $\theta = x_2$, we find
 \beq
ds^2 = \frac{-3/\Lambda}{(x_3 -x_{30})^2} \left( -dt^2 +dx_1^2 +dx_2^2 +dx_3^2 \right),
 \eeq
which is the $M_3$ foliation of pure $AdS_4$ manifold.
To conclude, even if we considered a more general class of warped solutions in the form of Eq.~(\ref{WarpM}), we found that there are no new solutions other than the known BTZ black string solutions.

\section{Solutions without warping factor}\label{solution}

In this section, let us consider the case that the geometry is not warped, {\it i.e.}, $H=1$ in Eq.~(\ref{WarpM}). Here we consider more general case in which the $\rho$ and $\theta$  coordinates are not necessarily radial and angular coordinates, respectively, {\it e.g.}, $\rho \to z$ and $\theta \to x$.
\begin{eqnarray}\label{withoutWarp}
ds^2= - f_0(z) dt^2 + f(z)dz^2 + f_1(z) dx^2+ f_2(z) dy^2 .
\end{eqnarray}
Note that these spacetimes can be regarded as a uniform foliation of three-dimensional spacetimes along the $y$ direction. Without loss of generality, we may set  $f(z)=f_0(z) f_1(z) f_2(z)$ by redefining the $z$ coordinate.
Thus, the most general form of the metric is given by:\footnote{This choice of metric functions makes the field equation, $R_{ab}- \Lambda g_{ab}=0$, be decoupled.}
\begin{eqnarray} 
ds^2=- e^{2U}(dt^2-e^{2V} dz^2)+  e^{2(V-W)} dx^2 + e^{2W} dy^2,
\end{eqnarray}
where $U$, $V$ and $W$ are functions of $z$ only. Solving the Einstein equation gives
\beqa
\label{UVW}
e^{2U} &=&\frac{\lambda}{\Lambda} \left(\frac{d}{12\lambda}\right)^{\frac13}
\left(\frac{2c}{1+d ~e^{-cz}}\right)^{\frac23} e^{-\frac13(2a+c) z},
\nn\\
e^{2V}&=&\left(\frac{d}{12\lambda}\right)^{\frac23}
\left(\frac{2c}{1+d~ e^{-cz}}\right)^{\frac43}e^{\frac23(a-c) z},\nn\\
e^{2W}&=&\left(\frac{d}{12\lambda}\right)^{\frac13}
\left(\frac{2c}{1+d ~e^{-cz}}\right)^{\frac23}e^{\frac13(b+a-c) z} ,
\eeqa
with a constraint equation
\beqa
a^2+\frac{b^2}{3}&=&c^2,
\eeqa
where $a, b, c,d$ and $\lambda$ are integration constants. This solution was found in Ref.~\cite{Wiltshire}. Note that, when $c=0$ and so $a=b=0$, the solution reduces to the pure $AdS_4$ or $dS_4$ metric depending on the sign of $\Lambda$, respectively. Henceforth, we consider  $c\neq 0$ case. However, the solution for $c=0$ can be obtained as well by taking the $c\rightarrow 0$ limit in the solutions with $c\neq 0$.

Let us introduce a new coordinate defined as
\beqa
\xi= \left( \frac{2c}{1+d ~e^{-cz}} \right)^{1/3} ~~\mbox{or}~~~
e^{-cz}= \frac{1}{d}\left(\frac{2c}{\xi^3}-1\right).
\eeqa
By suitable rescalings~\footnote{
$t\rightarrow|\Lambda|\left(2|d|^p/(3|\lambda|)\right)^{1/3}t,~~~~
x\rightarrow\left(12|\lambda|/|d|^{p-\sqrt3 q}\right)^{1/6}x,~~~~
y\rightarrow\left(12|\lambda|/|d|^{p+\sqrt3 q}\right)^{1/6}y.$}
of the coordinates $t$, $x$ and $y$, the metric in Eq.~(\ref{UVW}) can be rewritten as
\begin{eqnarray}
\label{rmetric}
ds^2&=&
-\frac{(-\Lambda) \xi^2}{3}\left(1-\frac{K^3}{\xi^3}\right) \left|1-\frac{K^3}{\xi^3}\right|^{\frac{2(p-1)}{3}}dt^2
+ \frac{d\xi^2}{\frac{(-\Lambda) \xi^2}{3}\left(1 -\frac{K^3}{\xi^3}\right)} \nn \\
&&~~~~~+ \xi^2
\left|1-\frac{K^3}{\xi^3}\right|^{\frac{1-p+q}{3}}dx^2 +\xi^2\left|1-\frac{K^3}{\xi^3}\right|^{\frac{1-p-q}{3}}dy^2.
\end{eqnarray}
Here  $K=\sqrt[3]{2c}$, $p=a/c$ and $q=b/c$, hence
\begin{equation}
p^2+\frac{q^2}{3}=1. \label{pq:constraint}
\end{equation}
Note that these solutions are characterized by two integral constants $K$, $p$ and the sign of $q$. We point out that the metric in Eq.~(\ref{rmetric}) is valid for both signs of $\Lambda$.

For $\Lambda <0$, the spacetime is asymptotically anti-de Sitter ($\xi \gg K$) and has a static boundary, allowing us to define ADM quantities. By using the Hamiltonian formalism in Refs.~\cite{Hawking:1995fd, Harmark:2004ch}, the gravitational mass and tension densities~\footnote{We point out that the ADM densities here are defined for unit coordinate intervals, not for unit proper distances, as in Refs.~\cite{Hawking:1995fd, Harmark:2004ch}.} associated with $t$, $x$ and $y$ translation symmetries  for stationary solutions can be obtained as
\beqa
\label{ADM:pq}
M \equiv\frac{(-\Lambda) K^3}{24\pi G}p  ,\quad
\tau_1\equiv-\frac{(-\Lambda) K^3}{48\pi G}(p-q),\quad
\tau_2\equiv-\frac{(-\Lambda) K^3}{48\pi G}(p+q) ,
\eeqa
respectively. We can see that the solutions denoted by Eq.~(\ref{rmetric}) are characterized by these three ADM parameters restricted on the plane given by
\beqa
\label{misteriousone}
M+\tau_1+\tau_2 =0.
\eeqa
Furthermore, actual values of these ADM parameters will be more restricted due to physical conditions for them although such conditions are not well known for asymptotically AdS spacetimes. Note that, in Eq.~(\ref{rmetric}), the spacetime described by $-\infty<\xi\leq 0$ is the same as the spacetime described by $0\leq \xi<\infty$ with opposite sign of $K$.
Thus we restrict the range of $\xi$ to be nonnegative. Suppose $K$ is negative, {\it e.g.,} $\bar K=-K>0$. Then, Eq.~(\ref{ADM:pq}) shows that the corresponding ADM parameters are the same for both $(K, p, q)$ and $(-K,-p,-q)$. Although the metrics look different very much for these two cases, this invariance indicates that both cases actually describe the same geometry. In fact, this equivalence can be explicitly shown by the coordinate transformation $\xi^3+\bar K^3\to \xi^3$ with $K\to \bar K$.
Note that we may restrict the value of $K$ be nonnegative using such symmetry.
For given values of $M$, $\tau_1$ and $\tau_2$, the integral constants appeared in Eq.~(\ref{ADM:pq}) can be obtained by
\begin{eqnarray}
K^3 =\frac{24  \pi G}{(-\Lambda)} \sqrt{\frac23(M^2+\tau_1^2+\tau_2^2)}, ~~
p = \frac{M}{\sqrt{\frac23(M^2+\tau_1^2+\tau_2^2)}}, ~~
q= \frac{\tau_1-\tau_2}{\sqrt{\frac23(M^2+\tau_1^2+\tau_2^2)}}.
\end{eqnarray}

The topology of the extra two dimensional space depends on how the exponents of $g_{xx}$ and $g_{yy}$ components behave. If $(1-p\pm q)/3 \neq 1$, the range of $x$ and $y$ could be unbounded and these two dimensional space are simply flat for a given $\xi$. Both directions or either one could also be compactified.
On the other hand, if $(1-p+ q)/3=1$ or $ (1-p- q)/3=1$, $x$ or $y$ should be an angle coordinate centered at $\xi=K$, respectively, describing a stringlike object rather than a two brane object. In this case the solutions are described by the mass density and a single tension.

The case of $p=1$ ({\it i.e.,} $\tau_1=\tau_2=-M/2$) with negative $\Lambda$ corresponds to the well-known solutions appeared in Refs.~\cite{Lemos:1994xs,Cai:1996eg,Horowitz,Huang:1995zb},
\begin{eqnarray}
\label{a=12}
ds^2 =-\frac{(-\Lambda)\xi^2}{3}\left(1-\frac{K^3}{\xi^3}\right) dt^2+\frac{d\xi^2}{\frac{(-\Lambda)\xi^2}{3}\left(1-\frac{K^3}{\xi^3}\right)}  + \xi^2(dx^2+dy^2).
\end{eqnarray}
This solution is a black 2-brane metric\footnote{As mentioned above, when one of the $x$ and $y$ coordinates  is compactified properly, this metric becomes the black string discovered by Lemos~\cite{Lemos:1994xs}.} in which the event horizon locates at $\xi=K$ and a spacelike curvature singularity exists at $\xi=0$.
The double Wick rotations of this solution for $t$ and $x$ coordinates corresponds to $p=-1/2$ and $q=\pm 3/2$ ({\it i.e.,} $(\tau_1,\tau_2)=(- 2M,M) \mbox{ or } (M, -2M)$), which maybe called as four-dimensional static bubble.~\footnote{This solution appeared as the so-called AdS soliton in Ref.~\cite{Horowitz}.} The metric for $q=3/2$ becomes
\begin{eqnarray}
\label{soliton}
ds^2&=&-\frac{(-\Lambda)}{3}\xi^2 \,dt^2 +
\frac{d\xi^2}{\frac{(-\Lambda)}{3}\xi^2 \left(1-\frac{K^3}{\xi^3}\right)} +\xi^2 \left(1-\frac{K^3}{\xi^3}\right)dx^2 +\xi^2 dy^2 .
\end{eqnarray}
Here, the conical singularity at $\xi=K$ can be removed by assigning a suitable periodicity on the coordinate $x$ ( {\it e.g.,} $x \equiv x+\frac{4\pi}{K\sqrt{-3\Lambda}}$). Thus, the range of the coordinate $\xi$ runs from $K$ to infinity, the geometry is regular everywhere, and its Penrose diagram is given in Fig.~{\ref{Pen:bubble}.
The geometrical properties for other cases will be analyzed in detail below.

\begin{figure}[thb]
  \begin{center}
  \includegraphics[width=.4\linewidth]{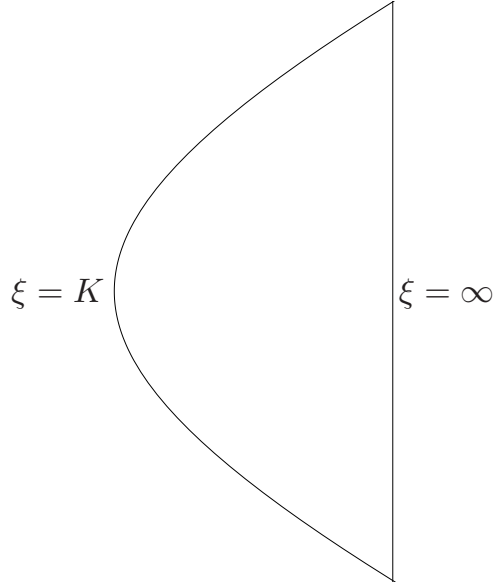}
  \end{center}
  \caption{Penrose diagram of the bubble solution
  }
\label{Pen:bubble}
\end{figure}

The curvature squared for the metric becomes
\begin{eqnarray}
\label{RR}
R^{abcd}R_{abcd}&=& 4\Lambda^2
    \left\{1+\frac{1}{6}(1+p)(2p-1)^2 \left(\frac{K^6}{\xi^6}
            -1\right)
      \right. \nn  \\
     &&~~~~~~~~~~\left. + \frac{1}{6}(1-p)(2p+1)^2
  \left[\left(\frac{\xi^3}{K^3}-1\right)^{-2}-1\right]
    \right\}.~~~~
\end{eqnarray}
Thus we see that curvature singularities occur at $\xi=0$ and $\xi=K$ generically.
However,  the curvature singularity exists only at $\xi=0$ for $p=1$ and at $ \xi=K$ for $p= -1$. These curvature singularities turn out to be mostly naked as will be explained in more detail below. The cases where the curvature singularity is enclosed by an event horizon are only for $p=1$ with negative cosmological constant.

\section{Properties of the solutions with negative cosmological constant }
\label{-L}

Let us consider the case of negative cosmological constant ($\Lambda<0$). It can be easily seen that all geometries in Eq.~(\ref{rmetric}) become pure AdS$_4$ as $\xi\to \infty$.
Thus the spatial infinity $\xi=\infty$ is the timelike AdS boundary.

For the case of $p=-1$, the metric becomes
\beqa
ds^2 &=& -\frac{(-\Lambda)\xi^2}{3}\left(1-\frac{K^3}{\xi^3}\right)\left|1-\frac{K^3}{\xi^3}
    \right|^{-4/3} dt^2 + \frac{d\xi^2}{\frac{-\Lambda \xi^2}{3}(1-K^3/\xi^3)} \nn \\
& & + \xi^2 \left|1-\frac{K^3}{\xi^3}\right|^{2/3}(dx^2+dy^2).
\eeqa
The Penrose diagram of this metric for the range $K\leq \xi < \infty$ is given in Fig.~\ref{Pen:-1}.
\begin{figure}[htb]
  \begin{center}
\includegraphics[width=.8\linewidth]{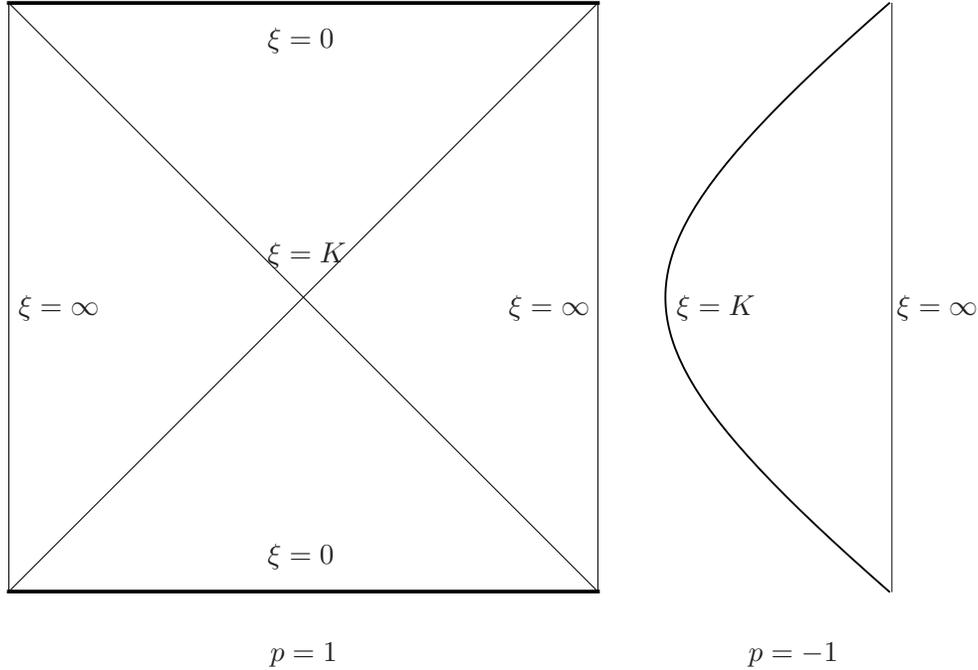}
  \end{center}
\caption{Penrose diagram of the solution with $p=\pm 1$. Thick lines denote curvature singularities.}
\label{Pen:-1}
\end{figure}
Namely, the curvature singularity at $\xi =K$ surface is naked. Note that $\xi$ plays the role of a time coordinate if $0< \xi < K$. Since $\xi=0$ surface is not a curvature singularity in this case, the spacetime will be extended even for negative $\xi$.
Allowing negative values of $\xi$, we can see that the spacetime covered by $-\infty<\xi\leq K$ is, in fact, equivalent to the case of $p=1$ with $0\leq \xi <\infty$.~\footnote{We use the change of variable $\xi^3-K^3=-{\xi '}^3$.}
Therefore, we do not need to consider the region with $\xi<K$ in this case.
We point out that the mass density of this solution in Eq.~(\ref{ADM:pq}) is negative and that two tension densities are equal and positive.

\begin{figure}[htb]
  \begin{center}
 \includegraphics[width=.5\linewidth]{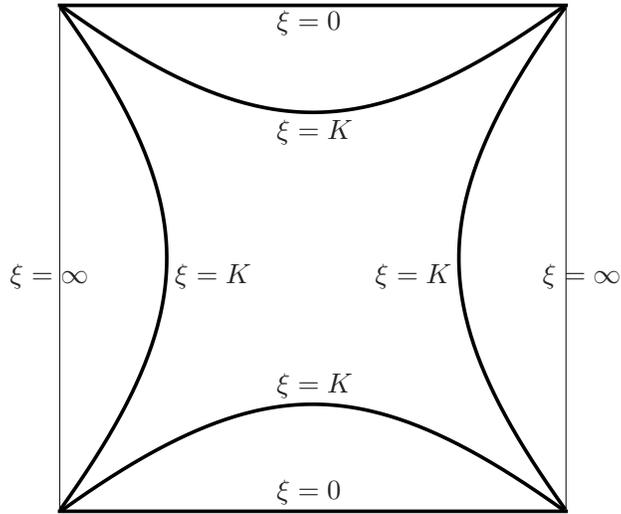}
  \end{center}
\caption{Penrose diagram of the solution with negative cosmological constant with $p\neq \pm 1,~\pm 1/2$. Thick lines denote curvature singularities.}
\label{fig:penrose_negative}
\end{figure}

For the cases other than $p= \pm1,~ \pm 1/2$,  there are  two curvature singularities at $\xi=0$ and $\xi=K$ as can be seen in Eq.~(\ref{RR}). The causal structure of this spacetime can be seen in Fig.~\ref{fig:penrose_negative}. The singularity at $\xi=K$ divides the solution into two disjoint regions, namely, a time-dependent one described by $\xi\in(0,K)$ and a static one by $\xi\in(K,\infty)$.

The static spacetime is asymptotically anti-de Sitter and bounded by a naked singularity inside at $\xi = K$. The geometry in $x$ and $y$ spatial directions is interesting as we approach to the $\xi =K$ surface. Whether its spatial distance shrinks or expands depends on the exponents in Eq.~(\ref{rmetric}). For a given mass density the sign of this exponents solely depends on the value of the tension density $\tau_i$. For instance, the spatial distance in $x$ shrinks to zero if $\tau_1>-M/2$ for positive mass density and if $\tau_1> M $ for negative mass density and {\it vice versa} for the spatial distance in $y$. For $\tau_1=-M/2(=\tau_2)$, both directions are regular. Note that $\tau_1+\tau_2~(=-M)$ should be conserved for a given $M$. For $M>0$, thus, if one of the spatial distance diverges, then the other should shrink as $\xi \to K$. For $M<0$, on the other hand, both distances shrink if $M<\tau_1<-2M$. The $x$ direction still shrinks, but $y$ direction diverges if $\tau_1>-2M$. If $\tau_1<M$, the $x$ direction diverges while the $y$ direction shrinks. For $\tau_1=-2M$ (and so $\tau_2=M$), the $x$ direction shrinks but the $y$ direction is regular.
These properties are similar to the case of hypercylindrical spacetimes in the absence of cosmological constant in five dimensions.

It is interesting to see the behaviors of spatial distances. At $t= {\rm constant}$ and $\xi= {\rm constant}$, the spatial lengths in $x$ and $y$ directions with unit coordinate distance are
\beq
L_x\equiv \sqrt{g_{xx}}=  \xi\left(1-\frac{K^3}{\xi^3} \right)^{\frac{1-p+q}{6}}
\quad {\rm and} \quad
L_y\equiv \sqrt{g_{yy}} = \xi\left(1-\frac{K^3}{\xi^3}\right)^{\frac{1-p-q}{6}},
\eeq
respectively. As $\xi \to K$, they shrink to zero or finite size, or diverge to infinity depending on the values of $p$ and $q$. The total volume becomes
\beq
V_{xy}\equiv L_x L_y = \xi^2 \left(1-\frac{K^3}{\xi^3} \right)^{\frac{1-p}{3}}.
\eeq
Thus we see that it always shrinks to zero as $\xi \to K$ except for the case of $p=1$ since $p\leq 1$ in Eq.~(\ref{pq:constraint}).

The time-dependent spacetime for $0\leq \xi <K$ is interesting. Since $\xi$ plays the role of time, we define a cosmological proper time as follows,
\beq
\bar t = \int_0^\xi\sqrt{g_{\xi'\xi'}}d\xi' = \frac{2}{\sqrt{3|\Lambda|}}\left[\frac\pi2- \tan^{-1}\left(\sqrt{\frac{K^3}{\xi^3}-1}\right)\right].
\eeq
Thus, $\xi=0$ and $K$ corresponds to ${\bar t}=0$ and $\pi/\sqrt{3|\Lambda|}$, respectively. In terms of the proper time and $t\to z$ the metric can be rewritten as
\begin{eqnarray}
ds^2 &=& -d{\bar t}^2 + K^2\left(\frac{\sin2\beta{\bar t}}{2\beta}\right)^{2/3}
    \left[\left(\frac{\tan\beta{\bar t}}{\beta}\right) ^{\frac{2(p-q)}{3}}d x^2 + \left(\frac{\tan \beta{\bar t}}{\beta}\right)^{\frac{2(q+p)}{3}} d y^2 \right.
     \nn \\
& & \qquad\qquad\qquad\qquad\qquad\qquad + \left.\left( \frac{\tan\beta{\bar t}}{\beta}\right)^{-\frac{4p}{3}}d z^2
    \right] .
\end{eqnarray}
Here $\beta = \sqrt{3|\Lambda|}/2$ and the coordinates $x$, $y$ and $z$ are scaled suitably. Note that this metric becomes exactly the Kasner solution as $\Lambda \to 0$~\cite{Kasner}. Therefore, this class of solutions we found is the extension of Kasner solution in the presence of negative cosmological constant. We point out that this AdS-Kasner solution starts from an initial singularity, but ends up with a final singularity, contrary to the case of Kasner solutions, in a finite time ($\pi/2\beta$).
Furthermore, the total spatial volume shrinks to zero at the future infinity due to the oscillating overall scale factor.

\section{Properties of the solutions with positive cosmological constant}
\label{+L}

In this section, we analyze the properties for the case of positive cosmological constant.
For $\Lambda > 0$, we reexpress the metric in~(\ref{rmetric}) as
\begin{eqnarray}
\label{rmetric:+}
ds^2 &=& -\frac{\Lambda \xi^2}{3}\left(\frac{K^3}{\xi^3}-1\right) \left|\frac{K^3}{\xi^3}-1\right|^{\frac{2(p-1)}{3}}dt^2
+ \frac{d\xi^2}{\frac{\Lambda \xi^2}{3}\left(\frac{K^3}{\xi^3}-1\right)} \nn \\
&&~~~~~+ \xi^2
\left|\frac{K^3}{\xi^3}-1\right|^{\frac{1-p+q}{3}}dx^2 +\xi^2\left|\frac{K^3}{\xi^3}-1\right|^{\frac{1-p-q}{3}}dy^2.
\end{eqnarray}
First of all, we see that the metric with $K=0$ corresponds to the pure de Sitter spacetime where the $\xi$-coordinate plays the role of time.~\footnote{$\xi = e^{\sqrt{\Lambda/3} \, \hat t}$ and $t=z/\sqrt{\Lambda/3}$ give the coordinate chart which covers only half the full de Sitter spacetime.} If $K\neq 0$,  we can set it to be $\pm 1$ by rescaling coordinates. As described below Eq.~(\ref{misteriousone}) for the case of $\Lambda<0$, we can also restrict the range of $\xi$ and $K$ as $0\leq \xi <\infty$ and $K\geq 0$, respectively.

The causal structures of the spacetimes in this case can be easily understood since the only difference from the case of  $\Lambda<0$ is that the signs of $g_{tt}$ and $g_{\xi\xi}$ are flipped. Consequently, the role of $t$ and $\xi$ coordinates are swapped and so the Penrose diagrams are same as those of the $\Lambda<0$ case, but tilted by $90$ degrees.

\begin{figure}[tb]
 \begin{center}
  \includegraphics[width=.8\linewidth]{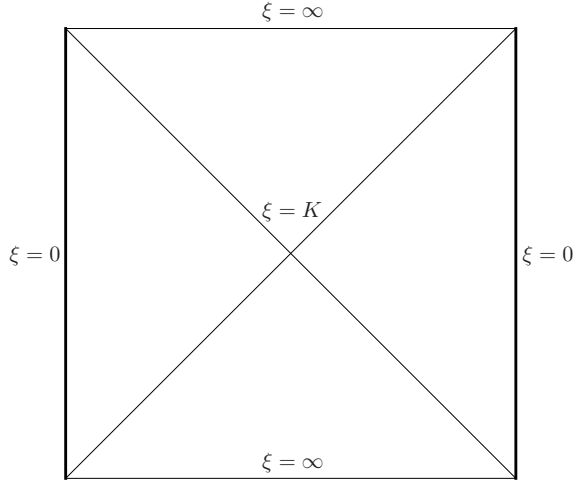}
 \end{center}
\caption{Penrose diagram of the solution with $p= 1,-1/2$ in the case of $\Lambda > 0$.}
\label{Pen:p1+}
\end{figure}

The case of $p=1$ is shown in Fig.~\ref{Pen:p1+}. The curvature singularity locates at $\xi =0$, which is naked. The spacetime region in $0 < \xi < K$ is static, but the region in $K < \xi < \infty$ becomes time-dependent, describing a Kasner spacetime. We point out that the spacetime with $p=-1/2$ is indeed the same as that with $p=1$. In the case of $\Lambda<0$, both metrics are related through the double Wick rotations, $\sqrt{-\Lambda/3}t\leftrightarrow ix$, while in the case of $\Lambda>0$ the two metrics are simply related by renaming $\sqrt{\Lambda/3} \,t \leftrightarrow x$.

\begin{figure}[thb]
  \begin{center}
 \includegraphics[width=.4\linewidth]{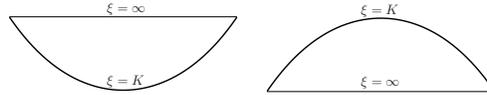}
  \end{center}
\caption{Penrose diagram of the solution with $p=- 1$ in the case of $\Lambda>0$.}
\label{Pen:p1-}
\end{figure}
The Penrose diagram for the case of $p=-1$ is given by Fig.~\ref{Pen:p1-}. The metric in the range of $\xi< K$ turns out to be equivalent to the metric with $p=1$ as in the case of $\Lambda <0$. The left hand side in the Penrose diagram describes a cosmological spacetime starting from an initial singularity and ending up with a de Sitter spacetime. The right hand side corresponds to a cosmological spacetime starting from a de Sitter metric which collapses to a singularity in the future.

Finally, the causal structures for the cases other than $p= \pm 1, \pm 1/2$ can be seen in the Penrose diagram in Fig.~\ref{Pen:gen+}. Explicit form of the metric describing the $K\leq \xi <\infty$ is given by
\begin{eqnarray}
ds^2 &=& -d{\bar t}^2 + K^2\left(\frac{\sinh2\beta{\bar t}}{2\beta}\right)^{2/3}
    \left[\left(\frac{\tanh\beta{\bar t}}{\beta}\right) ^{\frac{2(p-q)}{3}}d x^2 + \left(\frac{\tanh \beta{\bar t}}{\beta}\right)^{\frac{2(q+p)}{3}} d y^2 \right. \nn \\
& & \qquad\qquad\qquad\qquad\qquad\qquad  + \left.\left( \frac{\tanh\beta{\bar t}}{\beta}\right)^{-\frac{4p}{3}}d z^2
    \right] .
\end{eqnarray}
Here $\beta = \sqrt{3\Lambda}/2$ and the coordinates $x$, $y$ and $z$ are scaled suitably.
The comoving time ${\bar t}$ is related to the $\xi$ coordinate by
\beq
{\bar t} = \int^{\xi}_{K} \sqrt{|g_{\xi'\xi'}|}d\xi'
    =\frac1{\beta}\log \left(\sqrt{\frac{\xi^3}{K^3}}+\sqrt{\frac{\xi^3}{K^3}-1}\right).
\eeq
This class of solutions is indeed the extension of the Kasner solution in the presence of positive cosmological constant.

\begin{figure}[thb]
  \begin{center}
 \includegraphics[width=.8\linewidth]{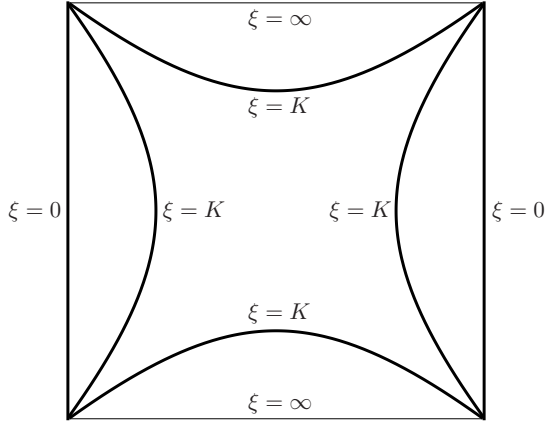}
  \end{center}
\caption{Penrose diagram of the solution with $p\neq \pm 1,~\pm 1/2$ in the case of $\Lambda>0$.}
\label{Pen:gen+}
\end{figure}

\section{Summary and discussions}
\label{discussion}

We have investigated string or branelike solutions of four-dimensional vacuum Einstein equations in the presence of cosmological constant. In the case of warped stringlike solutions, we have shown that the BTZ black string is the only solution in the form of metrics given in Eq.~(\ref{WarpM}).~\footnote{We mention that a warped BTZ black 2-brane solution is recently found even in five-dimensional gravity with a Gauss-Bonnet term~\cite{CuadrosMelgar:2007jx}.} The general solutions for the metric form~(\ref{withoutWarp}) in the case of nonwarped stringlike configurations are given by Eq.~(\ref{rmetric}). When the cosmological constant is negative, these nonwarped stringlike configurations are characterized by ADM mass ($M$) and tension densities ($\tau_1,~\tau_2$) with a constraint of $\tau_1+\tau_2=-M$. The case of equal tension densities ({\it i.e.,} $\tau_1=\tau_2=-M/2$) turns out to be the well-known black 2-brane metric~(\ref{a=12}). The case that one of the tension densities is twice of the other with opposite sign ({\it e.g.,} $\tau_1=-\tau_2/2=M$) is the AdS soliton spacetime~(\ref{soliton}) whose geometry is regular everywhere. The cases other than these two possess a naked singularity. The time-dependent solutions shown in Fig.~\ref{fig:penrose_negative} can be regarded as the AdS extension of the well-known Kasner solutions.
For the positive cosmological constant, on the other hand, the time-dependent metric turns out to be the Kasner-de Sitter spacetime~\cite{Kasner-deSitter} and the time independent solutions possess naked singularities.

It is worth considering the stability of solutions we described. In the absence of the cosmological constant, it was shown~\cite{Harmark:2005pp} that both the black string and the Kaluza-Klein bubble solutions in spacetime dimensions higher than or equal to five are unstable due to the Gregory-Laflamme instability~\cite{GL}. One may expect that the black 2-brane metric in Eq.~(\ref{a=12}) is also unstable due to the Gregory-Laflamme instability occurring in generic brane or stringlike spacetimes. According to the Gubser-Mitra conjecture~\cite{gubser}, however, this expectation might be wrong. The conjecture states that a black string or brane spacetime is stable if and only if it is thermodynamically stable. Being considered as a thermodynamic system, the black 2-brane in Eq.~(\ref{a=12}) gives $M= (8\pi^2k_B^3 /3 G|\Lambda|^2\hbar^3) T^3$ where the black hole temperature
$T=(\hbar|\Lambda|/4\pi k_B)K$ is used. Thus the heat capacity for the black hole is always positive, indicating classical stability of this spacetime through the Gubser-Mitra conjecture. Indeed the classical linearized stability has been studied for this background spacetime, showing the absence of instability~\cite{Bak:2010ry}.
In addition, Chen, Schleich and Witt~\cite{chen} showed that the warped AdS black string is also stable.

The static bubble solution in Eq.~(\ref{soliton}) turns out to be stable as well. This solution is a special case of $p=2$ in Ref.~\cite{Horowitz}, where Horowitz and Myers proposed the so-called new positive energy theorem. According to it, this metric is the ground state among all spacetime configurations having the same boundary behaviors.
Thus this spacetime is presumably stable under perturbations. Actually, the similar solution in the case of $p=3$ in Ref.~\cite{Horowitz}, which is called as  the AdS soliton spacetime in five dimensions, is shown to be stable under small perturbations~\cite{OKL}.

We have seen that solutions for the class of metric form~(\ref{withoutWarp}) are mostly singular except for several regular solutions for specific values of parameters.
What the implications of this fact would be? One may simply discard all these singular solutions. Or, since these metrics are static vacuum solutions, they might be the end states of matter collapses. Note that the physical parameters characterizing the initial matter distribution in such processes could be arbitrary as long as they satisfy suitable physical conditions.
Therefore, the appearance of many naked singular solutions through such dynamical processes could be an example of violating the cosmic censorship conjecture.

On the other hand, however, there might be a third scenario. For the spacetime solutions having naked singularities at the present consideration, their stability behaviors are not known as yet explicitly. We speculate though that these four-dimensional spacetimes are unstable, except for the case that the singularities are located at the past (future) infinity. In the case of five dimensions without having cosmological constant, these singular solutions are known to be unstable under small perturbations~\cite{kang}. If this speculation is true, such instability probably drives the matter collapse to some stable states before they reach to singular states, avoiding a violation of the cosmic censorship conjecture. Actually, a recent numerical study showed that, although there exists an electrovacuum solution containing a naked singularity, dynamical collapses of charged scalar fields in $(2+1)$ anti-de Sitter background do not produce such a singular solution~\cite{Hwang:2011mn}.

In the presence of positive cosmological constant, it was shown that the Kasner-de Sitter spacetime is unstable because the tensor mode perturbations increase indefinitely at the initial anisotropic stage~\cite{Kasner-deSitter}. However, it was also shown that the spacetime with $p=1$ is stable under the perturbation~\cite{km,Pitrou}.

\begin{acknowledgments}
This work was supported by the Korea Research Foundation Grant funded by the Korean Government (MOEHRD) KRF-2008-314-C00063 and in part by APCTP Topical Research Program. GK was supported in part by the Mid-career Researcher Program through an NRF grant funded by the MEST (No. 2008-0061333) and a research grant from SUN Micro Systems at KISTI. HCK was supported in part by a grant from the Academic Research Program of Chungju National University in 2011.

\end{acknowledgments}

\end{document}